\documentclass{aa}  
\usepackage{hyperref}
\usepackage{graphicx}
\usepackage{txfonts}
\usepackage{lipsum}
\usepackage{subcaption}         
\usepackage{lscape}             
\usepackage{placeins}           

\usepackage{ulem,xcolor,tikz} 

\begin{document}

   \title{Hostless extragalactic transients in Fink}
   \subtitle{Results from the ELEPHANT pipeline}

\author{R. Durgesh\inst{1}\thanks{\email{rupesh.durgesh4@gmail.com}}
           \and
            P. J. Pessi\inst{2} 
           \and
           E.~E.~O. Ishida\inst{3}
            \and
           J. Peloton\inst{4}
           }

   \authorrunning{Durgesh et al.}

\institute{
Independent Researcher,  Ingolstadt, Germany 
\and 
Astrophysics Division, National Centre for Nuclear Research, Pasteura 7, 02-093 Warsaw, Poland 
\and 
Universit\'e Clermont Auvergne, CNRS/IN2P3, LPCA, F-63000 Clermont-Ferrand, France
\and 
Université Paris-Saclay, CNRS/IN2P3, IJCLab, Orsay, France
}

   \date{Received September 30, 20XX}

 
  \abstract
   {The ExtragaLactic alErt Pipeline for Hostless AstroNomical Transients (ELEPHANT), has been developed as a framework for filtering hostless candidates, in real time alert systems, and implemented as a filter in the Fink broker. ELEPHANT works on stamps and requires minimal information, thus allowing for fast identification of extragalactic transient events. 
   }
   {In this work we evaluate the performance of the ELEPHANT pipeline by systematically analyzing flagged hostless candidates identified between 1 September 2023 and 31 December 2025. Our goal is to quantifying the pipeline's accuracy and identify dominant sources of contamination.} 
   {For each flagged candidate we collected additional information from multiple catalogues and archival repositories. We further examined their light-curve evolution and astrometric consistency (coordinate dispersion over time) to refine source classification.}
   {Out of 877 flagged events ($\sim$ 0.6\% of the total reported ZTF events in the analyzed period), 67 are confidently confirmed as genuinely hostless candidates, with no detectable host galaxy in either existing catalogues or archival imaging, representing a high-purity sample of intrinsically faint or absent hosts. Additional 51 events are linked to visually identifiable hosts that are entirely absent from both catalogues and ZTF stamps. 
   For the confirmed hostless subset, the inferred upper limits on host-galaxy absolute magnitudes ($M_r^{\mathrm{upper}} \approx -9.6$ to $-18$) extend well below the luminosity range of typical dwarf galaxies. The pipeline showed an overall accuracy of 84.03\%, with the majority of the classified flagged events being Type Ia supernovae, and the second most detected class being Type I superluminous supernova. ELEPHANT has been adapted to deal with the Rubin alert stream and has been processing its alerts since February 2026. 
  }
   {ELEPHANT has proven to be an efficient tool to automatically identify hostless transients in ZTF. Moreover, the framework proved to be adaptable to deal with the Rubin alert stream, thus enabling the identification of a large number of such candidates, paving the way for population studies of hostless events to an unprecedented limiting magnitude.}


   \maketitle

\section{Introduction}

The systematic study of local environments hosting extragalactic transients can provide important clues about the astrophysical processes, progenitor systems and general conditions necessary to trigger such events \citep{BERGER20111, Anderson2015}. Therefore, significant part of our current knowledge regarding such transients are direct consequences of such environmental studies \citep{Qin_2022}. 

In this context, it might not come as a complete surprise that the existence of apparent hostless extragalactic transients sparkles special interest within the astronomical community. These are events which cannot be directly associated to an underlying host or identifiable environment, thus raising multiple questions regarding the causality chains that might lead to their detection \citep{Strolger_2025}. It is expected that between 2\% to 7\% of observed transients can be identified as hostless \citep{qin2024statisticsenvironmentshostlesssupernovae}.  

Two main scenarios are considered for the non-detection of expected hosts. The most probable cause being the limiting magnitude of currently available observational facilities.  Faint dwarf galaxies, that lie below the survey's detection limit, remain undetected unless a bright transient occurs and outshines its host galaxy. In this context, the transient serves as a probe for discovering faint, hidden galaxies and offers a window into their obscured environments \citep{2012A&A...538A..30Z}. On the other hand, they may be associated with supernovae from escaped hypervelocity stars,  whose progenitor moves fast enough to escape their host after dynamical interactions with another stellar system \citep{2011A&A...536A.103Z, Perets_2012, Przybilla_2008}. Although extremely interesting, this scenario is very unusual, particularly for core-collapse supernovae because of their short life spans and time required to travel long distances \citep{Strolger_2025}. Other possible hypothesis include supernovae from stars in the tidal tails and diffuse stellar systems resulting from galaxy mergers  \citep{2015ApJ...807...83G} and supernovae originating from intracluster stellar systems \citep{gal-yam2003}. Finding  apparently hostless transients provides an unique opportunity to explore such scenarios. 

Moreover, systematic searches for hostless events can increase the number of detected faint, low surface brightness, and/or dwarf galaxies \citep{Conroy_2015,qin2024statisticsenvironmentshostlesssupernovae}, essential for constraining the faint-end of the galaxy luminosity function and understanding galaxy formation, evolution, and the $\Lambda$CDM model \citep{sales2022}. Moreover, it allows us to investigate the star formation history of galaxies, particularly in dwarf galaxies, and to study their star formation rates and mass-metallicity relations \citep{Strolger_2025}. 

Regardless of their significant scientific potential, the observational challenges associated with assembling a large enough sample of hostless transients to enable population studies has only recently being surpassed. The systematic observation of large areas of the sky, currently being carried by surveys like the Zwicky Transient Facility \citep[ZTF,][]{Bellm_2019}, and soon the Vera C. Rubin Observatory Legacy Survey of Space and Time \citep[LSST,][]{LSST_2009}, present a great opportunity to produce large, statistically significant samples using automated filtering techniques. 

In this context,  the ExtragaLactic alErt Pipeline for Hostless AstroNomical Transients\footnote{\label{fn:pipeline}\url{https://github.com/astrolabsoftware/fink-science/tree/master/fink_science/ztf/hostless_detection}} \citep[ELEPHANT,][]{Pessi_2024} was developed to enable real time processing of ZTF data. It consists of a statistical pipeline, implemented within the Fink broker\footnote{\url{https://fink-broker.org/}} \citep{Moller_2021},  which automatically identifies extragalactic hostless transients and redirects the most promising ones to be further scrutinized by experts. The pipeline has been processing ZTF alerts since August 2024, identifying on average 3 alerts per night. 

In this work, we analyze the performance of ELEPHANT, summarize the characteristics of reported candidates and  assess its contamination rate. Moreover, we describe  the  changes already made to the pipeline implemented in Fink, required to process the LSST stream\footnote{\label{fn:lsst_pipeline}\url{https://github.com/astrolabsoftware/fink-science/blob/master/fink_science/rubin/hostless_detection/processor.py}}. This paper is structured as follows: Section \ref{sec:pipeline_recap} details the ELEPHANT pipeline. Section \ref{sec:data} describes the dataset used in this work. Section \ref{sec:analysis} presents the cross-matching and contamination rate analysis. In section \ref{sec:elephant_for_lsst} we explain the modifications in the pipeline to process LSST alerts. Finally, Section \ref{sec:conclusions} presents our conclusions.

\section{The ELEPHANT pipeline}
\label{sec:pipeline_recap}

\begin{figure}
    \centering
    \includegraphics[scale=0.3, angle=-90]{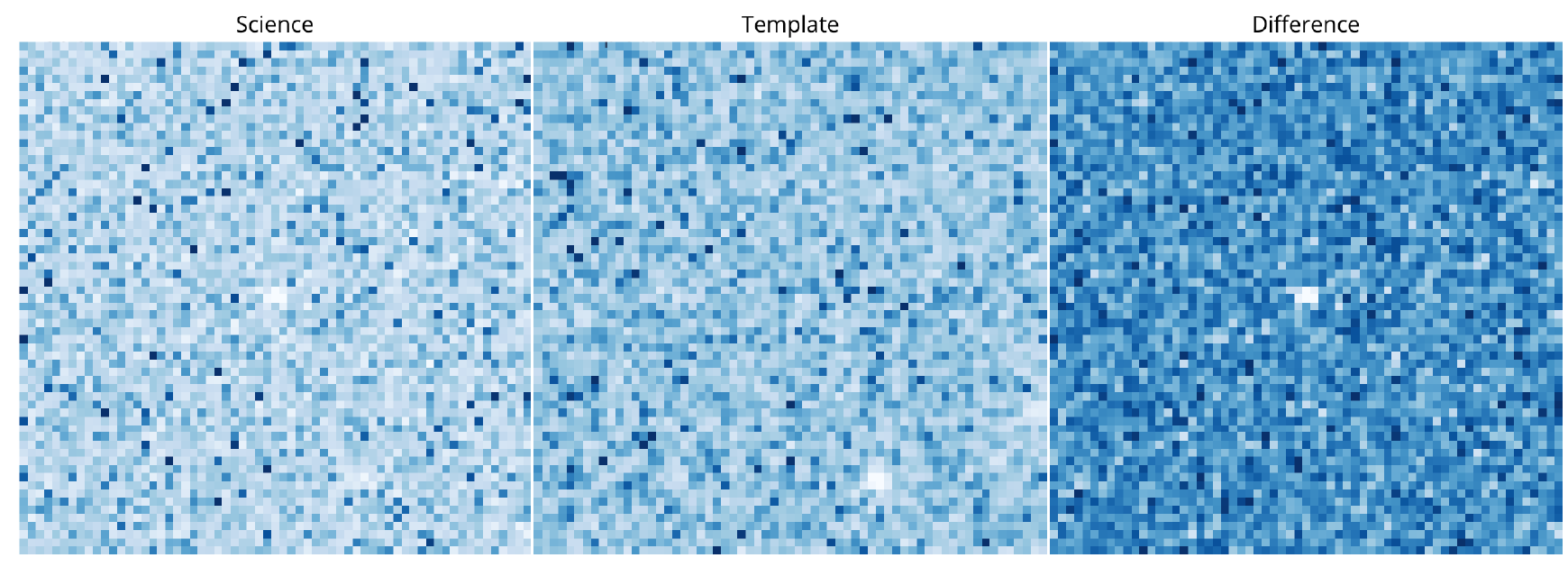}
\caption{SN 2025kkb / ZTF25aaofttb\protect\footnotemark: Example of hostless template stamp.}
 \label{fig:hostless_stamp_example}
\end{figure}

ELEPHANT takes science and template stamps from the ZTF alerts as inputs (Figure \ref{fig:hostless_stamp_example}). It starts by creating cropped images in the center of the stamp  (15 $\times$ 15 pixels, corresponding to $\sim$ 15 x 15 arcsec) for both, science and template images. For each cropped image, sigma clipping is applied ($\sigma$ = 3) to detect pixels potentially associated to a bright source. If bright pixels are found in the science, but not in the template image, or vice versa, the candidate is selected as potentially hostless and submitted to the power spectrum analysis.

In the subsequent stage, it assumes that an original image with a faint host encodes the presence of the host in the statistical properties of its pixel distribution in frequency domain, even if not identified by the sigma clipping step. Thus, the image is intrinsically  distinguishable from an image created by randomly shuffling noisy pixel values, as this shuffled image would not retain spatially coherent information. The pipeline transforms the pixel values into the frequency domain and then computes the Wasserstein distance between the original image and a set of its shuffled counterparts. Finally, it  uses a Kolmogorov–Smirnov (KS) test to compare these two distributions. If the distributions differ significantly, the original image is considered to contain a source. The KS-test thresholds were determined by running the pipeline on a test set of images, with known labels (hostless vs. containing structure), obtained through visual inspection. 
ELEPHANT has been integrated as a hostless detection science module in the Fink alert broker \citep{Moller_2021} in August 2024, associated with a dedicated public  bot\footnote{\label{fn:elephant_telegram}\url{https://t.me/fink_hostless}}, which daily delivers hostless candidates to the community.

The pipeline parameters have been empirically calibrated since its integration into the broker. The live pipeline processes extragalactic transient (see Appendix \ref{app:fink_filters}) alerts that have been classified by various science modules within Fink. It only processes alerts discovered within the last 45 days. Since our goal is to better understand these events, we encourage spectroscopic observations of the reported targets. A magnitude cut was also applied to further restrict the reported sample to candidates suitable for follow-up by current available instruments. The latest detection should be brighter than 19.5 mag, which is a reasonable brightness for most spectroscopic facilities to obtain a spectrum of a young transient event. Furthermore, at least three detections are required to ensure that the alert corresponds to a real extragalactic source. Alerts related to asteroids are ignored using the \texttt{d:roid} column provided within the alert package.
The filtering configuration and updated pipeline can be found in the Fink science GitHub repository\footref{fn:pipeline}.

\footnotetext{\url{https://ztf.fink-portal.org/ZTF25aaofttb}}

The pipeline has been recently integrated into the Fink broker to also process LSST alerts (see Section ~\ref{sec:elephant_for_lsst}). This implementation is currently in its initial stages which require extended tests and validation by domain experts, as the survey is still in its initial phase. As it progresses, improved image templates and a larger alert sample will become available, enabling ongoing refinement of the pipeline parameters and steady improvement in performance.

\section{Data}
\label{sec:data}

ELEPHANT processes ZTF real-time alerts ingested by the Fink broker. Fink focuses on enhancing the scientific value of each alert through pre-processing, classification, and cross-matching, for which a series of statistical and machine learning approaches are used\footnote{\url{https://fink-broker.org/papers/}}. In this work, we used the Fink data transfer service\footnote{\label{fn:fink_datat_transfer}\url{https://ztf.fink-portal.org/download}} to download ZTF alerts associated to extragalactic transients (see Appendix \ref{app:fink_filters}). This association is based on Fink classifications, obtained by cross-matching with SIMBAD \citep{wenger2000simbad} and the Transient Name Server (TNS\footnote{\label{fn:tns}\url{https://www.wis-tns.org/}}), or by using machine learning (ML) classifiers deployed within the broker. We do not use alerts that have been associated with Solar System objects or cataclysmic variable stars (CVs). Asteroid related alerts are excluded using the \texttt{d:roid} column within the alert package.

\begin{figure}
    \centering
    \includegraphics[width=0.5\textwidth]{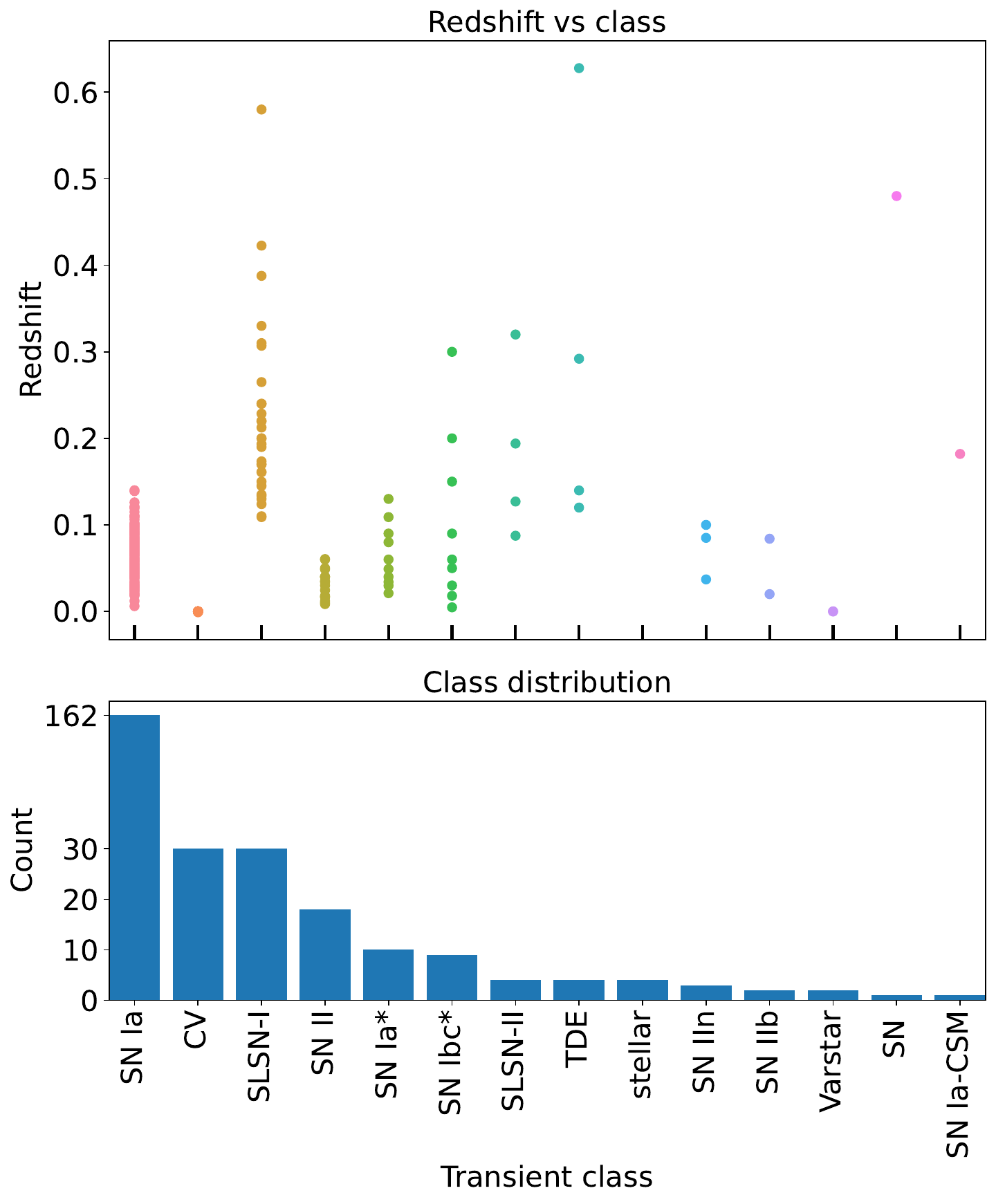}
\caption{Number of classified hostless candidates and their redshift distributions, showing a total of 276 spectroscopically confirmed sources in TNS. SN Ia-91bg-like, SN Ia-91T-like, SN Ia-pec, and SN Ia-SC are mapped to SN Ia*; SN Ib-Ca-rich, SN Ibn, SN Ic-BL, and SN Ib/c are mapped to SN Ibc*; SN II and SN IIP are mapped to SN II; TDE and TDE-featureless are mapped to TDE; CV and Nova are mapped to CV.}
    \label{fig:class_distribution_and_redshift}
\end{figure}

We downloaded alerts from 1 September 2023 to 31 December 2025, covering over two years of the ZTF stream. This corresponds to 3,193,902 alerts for 156,110 ZTF objects. The pipeline flagged 3,272 alerts from 877 ZTF objects as hostless candidates. Among these, 276 have spectroscopic classification available on the Transient Name Server\footnote{\url{https://www.wis-tns.org/}}. Their class and redshift distribution are shown in Figure \ref{fig:class_distribution_and_redshift}.
Among the classified objects, 162  ($\sim 58.69\%$) are Type Ia Supernovae (SN Ia) lying within the local universe ($z < 0.14$);  30 ($\sim 10.86\%$) are Superluminous Supernovae I (SLSN-I) in the redshift range of $0.1 < z < 0.58$; 
and  30 ($\sim10.86\%$)  have been classified as Cataclysmic Variable stars (CVs). The latter constitute the main source of contamination for our pipeline and will be further discussed in Sect.~\ref{sec:contaminationCVs}. We also identified  18 Type II Supernovae (SN II), 4 Superluminous Supernovae II (SLSN-II) and  4 TDEs, including a featureless TDE at redshift $z = 0.6278$ \citep{2025TNSAN.131}. Remaining 28 objects belong to additional classes, presented  in Figure \ref{fig:class_distribution_and_redshift}.

\begin{figure}
    \centering
    \includegraphics[width=0.5\textwidth]{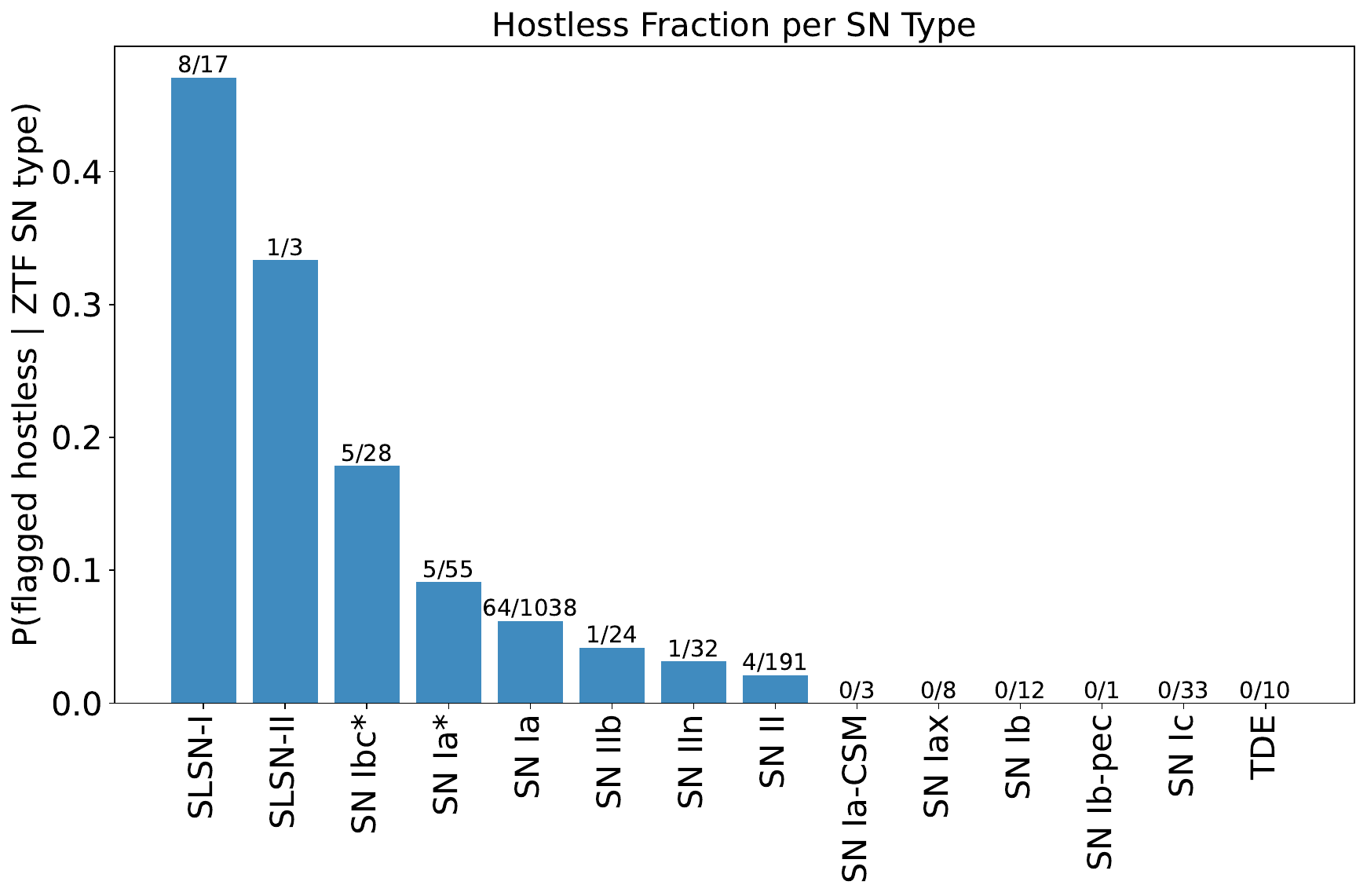}
\caption{Fraction of the BTS SN sample flagged as hostless by ELEPHANT, as a function of sub-class assigned to each transient. The distribution shows only candidates within the BTS magnitude limit of 18.5 mag that peaked between September 1st 2023 and December 31st 2025, and that pass quality and purity cuts. }
    \label{fig:hostless_fraction}
\end{figure}

The fraction of flagged hostless transients relative to the SN rates is shown in Figure \ref{fig:hostless_fraction}. We compute the fraction of hostless transients using ZTF Brightness Transient Survey \citep[BTS,][]{2020ApJ...895...32F, 2020ApJ...904...35P, 2024ApJ...972....7R} flux-limited sample data, which corresponds to sources with peak magnitude brighter than 18.5 along with pass quality and purity checks,  between 1 September 2023 and 31 December 2025. In this figure, we only included flagged hostless candidates that also appear in the ZTF BTS sample. Among the BTS SLSN‑I sample, $47\%$  percent  were flagged as hostless, followed by SLSN‑II at $33.3\%$. The BTS data are dominated by SN Ia, of which $6.1\%$ were flagged as hostless. The second most common class, SN II, has $2\%$ of its corresponding sources flagged as hostless by our pipeline.

These results are in agreement with \cite{qin2024statisticsenvironmentshostlesssupernovae}, which states that despite SLSN representing only a small fraction of the SN population \citep{2020ApJ...904...35P}, they account for a significant fraction of the detected hostless transients. This is a direct consequence of them being extremely energetic events whose brightness largely surpasses that of their local environments, thus acting like beacons for lower surface brightness and dwarf galaxies. 

\section{Analysis}
\label{sec:analysis}

The ELEPHANT pipeline was designed to operate with a minimal set of inputs, relying only on the alert stamps and the aggregated features provided by the Fink broker. Its goal being to lower the burden of visual inspection which the final analysis imposes in the domain expert. In this section, we further examine the different sources of contamination present among our flagged candidates (see Section ~\ref{sec:data}) and use them as a proxy to quantify the performance of the pipeline.

\subsection{Contaminants}
\label{sec:contaminants}

\subsubsection{Cataclysmic Variables}
\label{sec:contaminationCVs}

\begin{figure}
        \centering
        \includegraphics[width=1\linewidth]{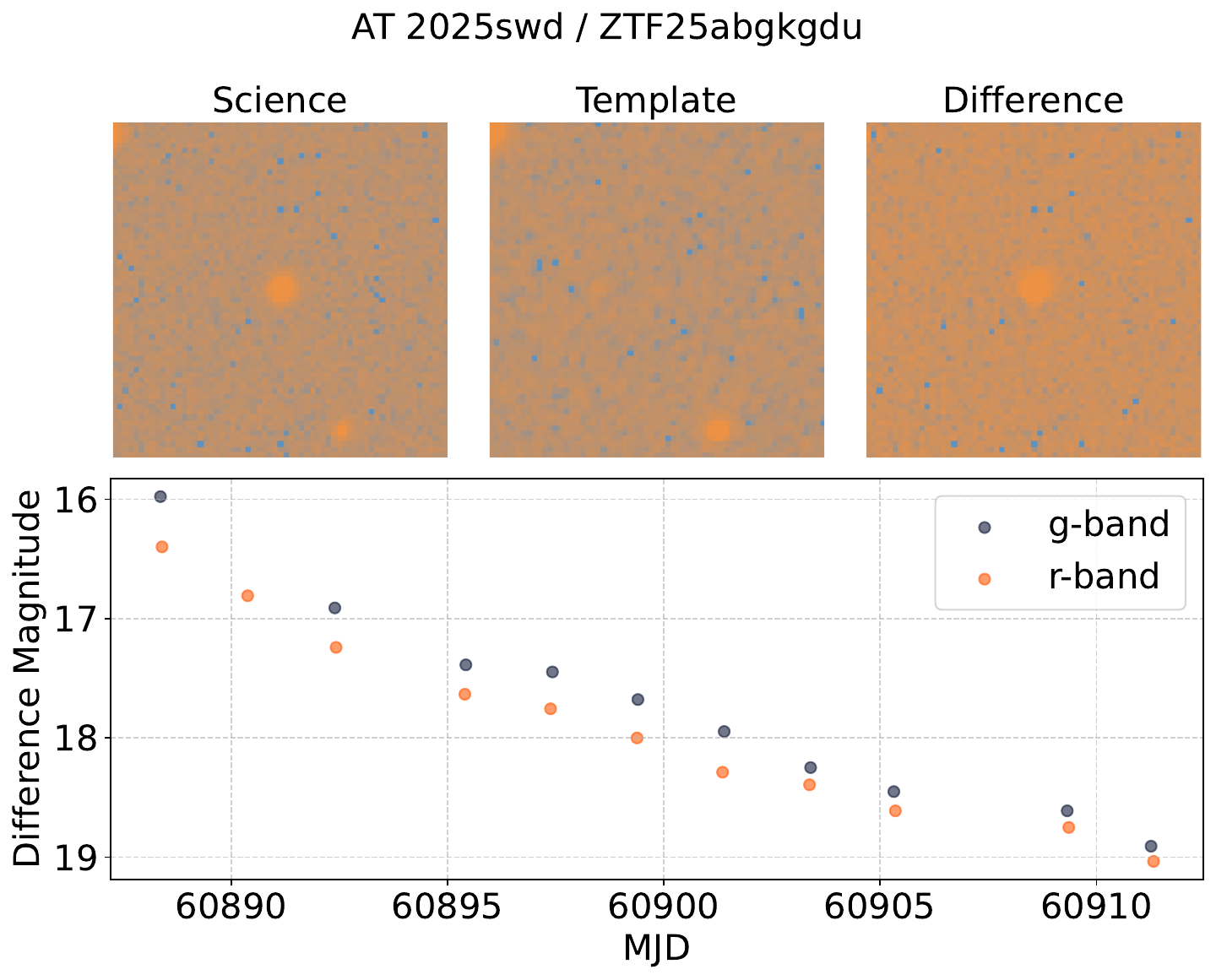}
      
    \caption{AT 2025swd (ZTF25abgkgdu), an example of a spectroscopically classified CV. This event is correctly identified as a hostless candidate, as it satisfies the selection criteria of the pipeline, which only considers  the central part of the stamp. However, it does not correspond to an extragalactic source. Top panel shows the science, template and difference stamps. The bottom panel displays its characteristic light curve, which is relatively bright, rapidly evolving, and shows an approximately linear decline.}
    \label{fig:ZTF25abgkgdu_CV_example}
\end{figure}

One of our main sources of contamination are CVs, in particular CVs that appear in isolated areas of the sky. The pipeline cannot determine whether these events are galactic or extragalactic \citep[see][for extragalactic novae and their rates]{2000NewAR..44...87S,2003A&A...405...23M,2020A&ARv..28....3D} and will correctly flagged them, as they are not associated to a visible host (see Fig.~\ref{fig:ZTF25abgkgdu_CV_example} for an example). 

In the case of hostless candidates without spectroscopic classification, we perform a visual inspection of their light curves and identify events as potential CVs when their evolution is consistent with typical CV-like light curves (bottom panel of Fig.~\ref{fig:ZTF25abgkgdu_CV_example}). In total, $15.84\%$ (139 of 877) of the flagged candidates are CVs. Among these, $10.86\%$ (30 of 276 transients on TNS) of the spectroscopically classified objects are CVs. 
The full list of CVs is available in our Github repository\footnote{\label{fn:supplementary_data_ref}\url{https://github.com/utthishtastro/extragalactic_hostless_analysis/tree/main/supplementary_data}}.

\subsubsection{Moving stars}

\begin{figure}
    \centering
    \includegraphics[width=0.5\textwidth]{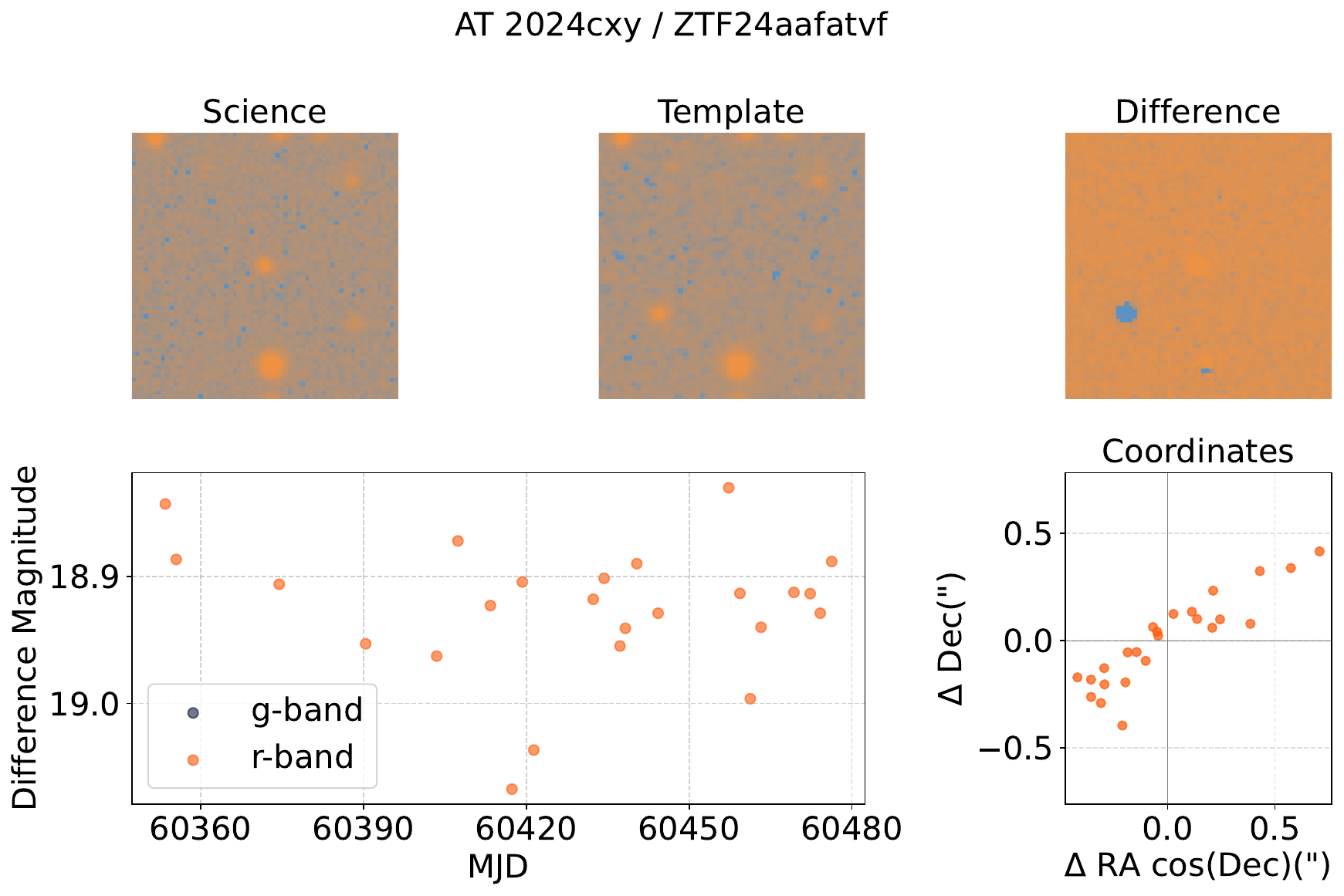}
    \includegraphics[width=0.5\textwidth]{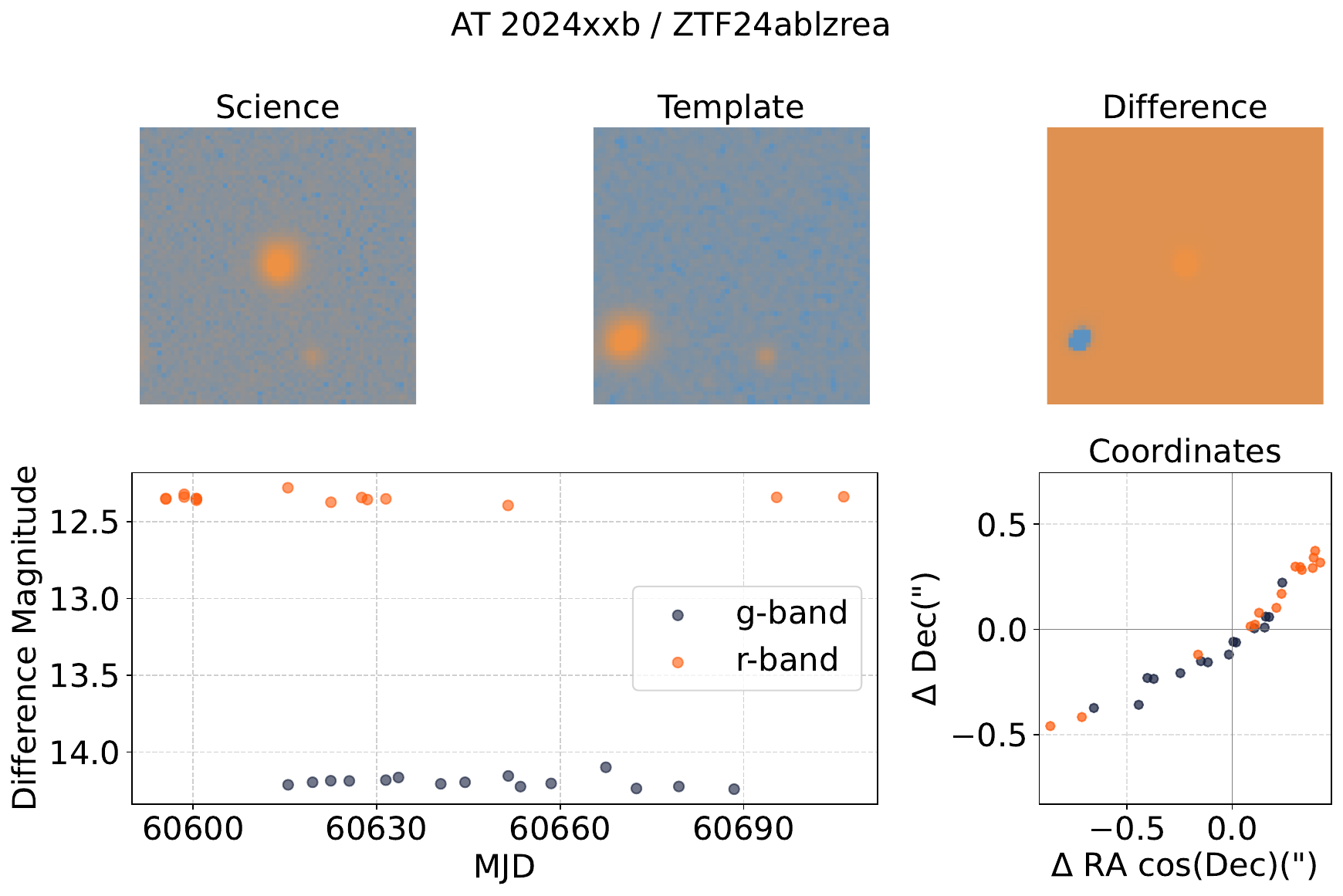}
    \includegraphics[width=0.5\textwidth]{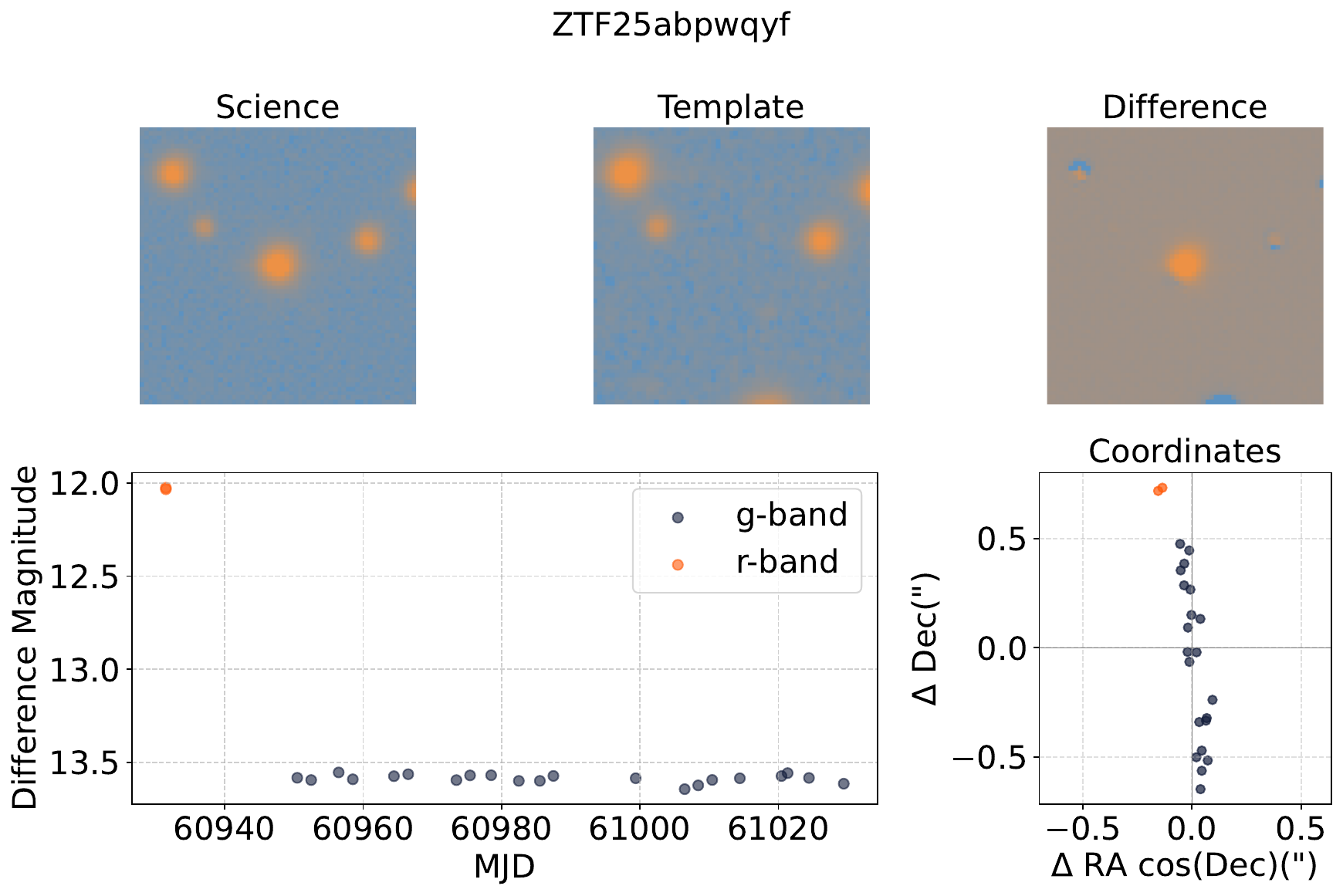}
    \caption{High proper motion stars identified within flagged hostless candidates. Alert stamps, light curve and coordinate dispersion as obtained from the Fink broker for SSSPM~J1444-2019 (top), Wolf~359 (middle) Ross 619 (bottom).}
    \label{fig:moving_stars}
\end{figure}

Another source of contamination are high proper motion stars in the Milky Way. Our pipeline has flagged 3 of them as hostless: (i) AT~2024cxy/ZTF24aafatvf\footnote{A similar transient light curve, designated AT 2023jjz/ZTF23aalfhif, was identified on the TNS within 5 arc seconds. However, the counterpart was not flagged because it lies outside the time frame of the data used in the paper.}
, corresponding to the ultracool subdwarf SSSPM J1444-2019 \citep{2004A&A...428L..25S}; (ii) AT~2024xxb/ZTF24ablzrea\footnote{A similar transient light curve, designated AT 2024dum/ZTF24aaeirya, was identified on the TNS within 5 arc seconds. However, the counterpart was not flagged because initial alerts had a Fink classification of Unknown, and later alerts did not meet the 45-day threshold.}, corresponding to the remarkably active M-dwarf star Wolf 359 \citep{1969ApL.....3..149G,piller1990,2025AJ....170..297L}; and (iii) ZTF25abpwqyf\footnote{ A similar transient light curve, designated AT 2025hkk/ZTF25aanbqyy, was identified on the TNS within 5 arcseconds. However, the counterpart was not processed because it has a Fink classification of Unknown, which is ignored by the pipeline (Appendix \ref{app:fink_filters})} corresponding to Ross 619 \citep{1927AJ.....37..193R}. Their stamps satisfy the conditions for the pipeline to identify them as hostless (see Fig.~\ref{fig:moving_stars}) but, as in the case of CVs, they do not correspond to extragalactic classes targetted in this work. We can identify these events as contaminants based on their characteristically flat  light curves and by the observed trend between coordinates of individual alerts (see Figure~\ref{fig:moving_stars}). By considering CVs and moving stars as contaminants, our pipeline achieves an overall accuracy of 84.03\%.

\subsection{Catalog cross-matching}
\label{sec:catalog_cross_match}

In order to further investigate the characteristics of the proposed candidates without a spectroscopic confirmation, we augmented the cross-matching available in Fink with large external data bases. We considered Pröst \citep[][Section \ref{subsec:prost}]{Gagliano2025_Prost}, a Python package developed for transient host galaxy association, and the Sherlock \citep[][Section \ref{subsec:sherlock}]{Young2023} system for contextual transient classification.

\subsubsection{Cross-matching with Pröst}
\label{subsec:prost}

Pröst can query major catalogs, such as GLADE+ \citep[Galaxy List for the Advanced Detector Era,][]{Daly2022_GLADEplus}, DECaLS \citep[Dark Energy Camera Legacy Survey DR9 and DR10,][]{Dey2019_DESI}, Pan-STARRS \citep[The Panoramic Survey Telescope and Rapid Response System DR1 and DR2,][]{Chambers2016_PS1}, and SkyMapper \citep[DR4,][]{Onken2024_SkyMapperDR4}, and uses Bayesian methods to estimate posterior probabilities to determine the likelihood of a candidate galaxy being the true host of a transient. It works by combining prior information about the host properties (fractional radial offset, absolute magnitude, and optionally redshift), with likelihood functions that quantify the goodness-of-fit between each candidate and the transient's observed properties. The posterior probabilities provide a ranked list of potential hosts.

We initialize priors and likelihood functions following a similar approach to that of Blast \citep{Jones2024_BLAST}, a framework for characterizing transients' host galaxies. The priors are as follows: an offset is assumed to be uniformly distributed within a radius of 5 kpc from the transient's location; absolute magnitudes are modeled with a uniform distribution spanning the range of $-$30 to $-$10;  a half-normal distribution is used for the redshift prior, with a mean of 0.0001 and a scale parameter of 0.5,  which is only considered when the transient's redshift is known.
The gamma (a=0.75) and SnRateAbsmag (a=-25, b=20) likelihoods are used for offset and absolute magnitude, respectively.

Among the 877 flagged hostless candidates (including those with spectroscopically classified transients), Pröst identifies potential host galaxies for 672 events. Of these, 377 are associated with a Pan-STARRS source, 292 with a DECaLS source, and 3 with a GLADE+ catalog source. Host associations were found for 230 of the 276 candidates with spectroscopic classifications (Fig.~\ref{fig:host_association_count}, top panel).

\subsubsection{Cross-matching with Sherlock}
\label{subsec:sherlock}

Sherlock supports many large sky surveys such as Gaia DR1 and DR2 \citep{GaiaCollaboration2016, GaiaCollaboration2018}, Pan-STARRS,  SDSS DR12 PhotoObjAll and SDSS DR12 SpecObjAll Tables \citep{Alam2015}, 2MASS point- and extended-source catalogues \citep{Skrutskie2006} and Guide Star Catalogue v2.3 \citep{Lasker2008}. It associates hosts to transients by calculating the angular separation between a transient's position and catalogued galaxies in its surroundings. An association is rejected if morphological measurements are available for the galaxy and the angular separation exceeds 2.4 times its semi-major axis. If redshift or distance information is available for the galaxy, angular separations are converted to projected physical distances, and associations with projected separations greater than 50 kpc are rejected.

\begin{figure}
    \centering
    \includegraphics[width=0.5\textwidth]{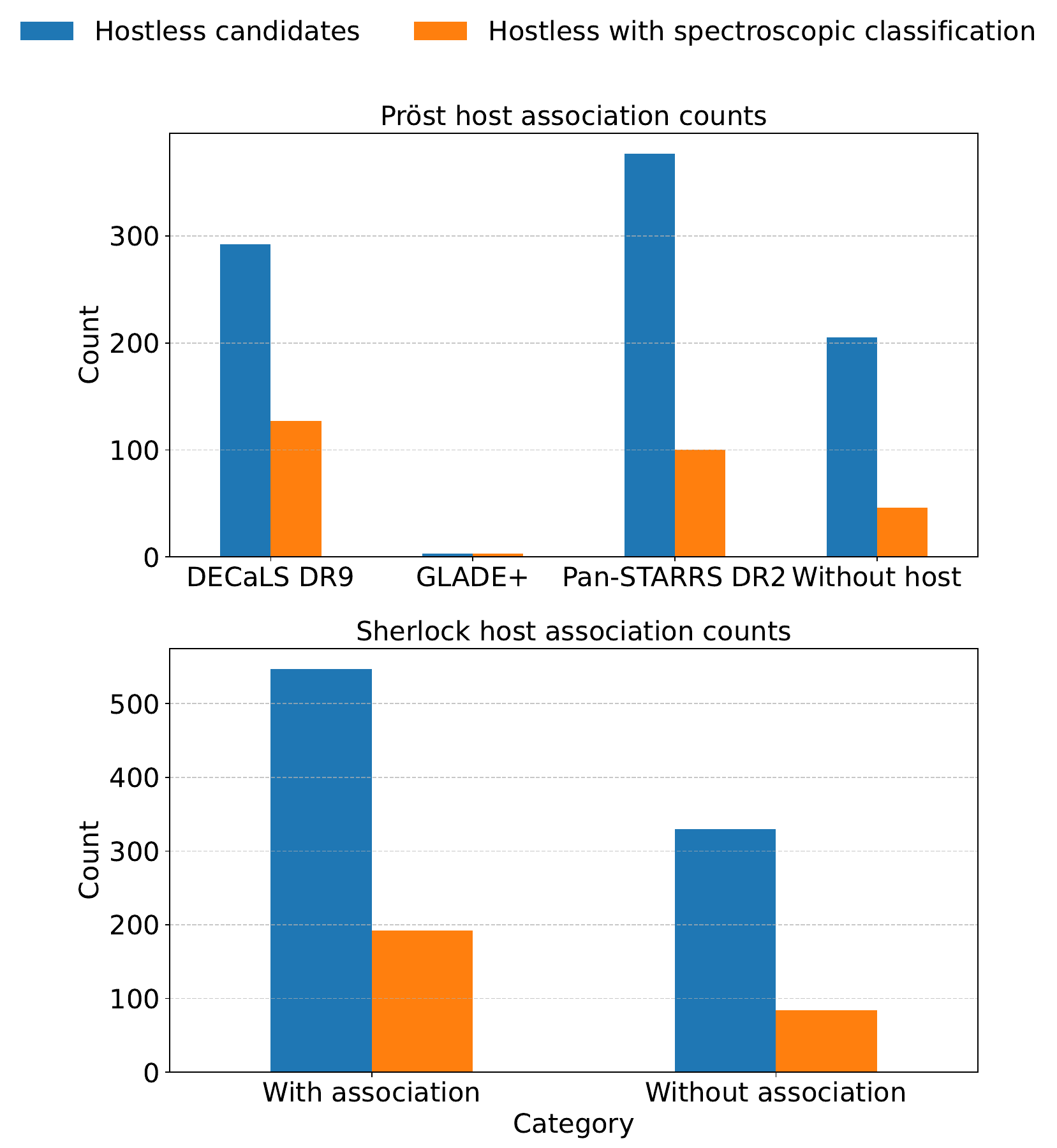}
    \caption{Pröst (top) and Sherlock (bottom) host association counts. In blue we show all the hostless candidates with a host association, while the candidates with spectroscopic classification on TNS are shown in orange. Pröst associations are divided by catalogue, Sherlock associations only consider the presence or absence of a host, regardless of the catalogue where the host was found.}
    \label{fig:host_association_count}
\end{figure}

Sherlock finds associated host galaxies from different catalogues for 547 of the 877 flagged hostless candidates. Host associations were found for 192 of the 276 hostless candidates with spectroscopic classifications (Fig.~\ref{fig:host_association_count}, bottom panel).

\subsubsection{Cross-matching comparison}

To compare host assignment between Sherlock and Pröst, we checked the frequency at which both packages assigned identical hosts from the PS1 catalog. We found that 61.72\% of PS1 cross-matched objects had consistent assignments in both packages. Next, we evaluated transients with cross-matching in both catalogs and compared their reported r-band magnitudes (shown in Fig.~\ref{fig:host_magnitude_distribution}). In 75\% of cases the magnitude difference was $\leq0.5$~mag; in the remaining 25\%, the differences exceeded 0.5~mag. Although cross-catalogue magnitude discrepancies are expected for extended sources due to differences in aperture definitions, source deblending algorithms, and sky subtraction strategies, differences larger than 0.5~mag exceed the median scatter reported in cross-survey comparisons \citep[][]{2019MNRAS.490..634S}, indicating possible discrepancies in the identification of host galaxy between the two frameworks. While, in this work, the presence of a catalogue-associated host is sufficient to discard an event as a hostless candidate, such disagreements between catalogues suggest that some associations may be uncertain. Consequently, a subset of these cases could still be consistent with genuinely hostless candidates. A detailed assessment of the accuracy of host-galaxy association is beyond the scope of this work.

\subsection{Host magnitude distribution}

\begin{figure}
    \centering
    \includegraphics[width=0.5\textwidth]{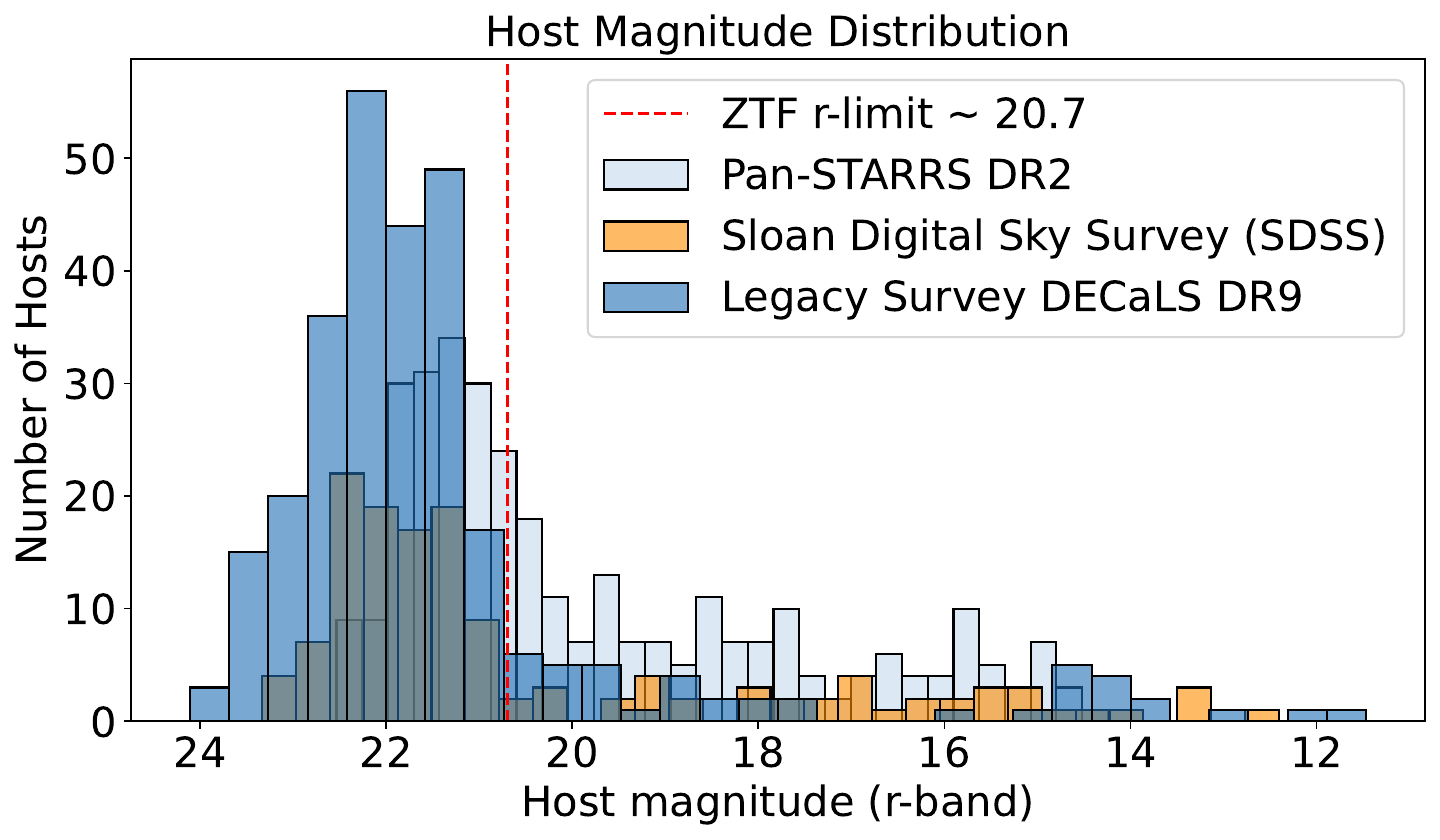}
    \caption{Catalogued host $r$-band magnitude distribution. The ZTF $r$-band magnitude limit is shown as a dashed vertical red line.}
    \label{fig:host_magnitude_distribution}
\end{figure}

The lack of host detection can be associated to the depth of the survey. In this section we compare the host magnitudes reported in the corresponding catalogues (see Section ~\ref{sec:catalog_cross_match}) to the ZTF limiting magnitude. We consider the \textit{r}-band, as it tends to be better observed and is less affected by potential reddening. The ZTF \textit{r}-band $5\sigma$ limiting magnitude for a single exposure (the limit for detecting a faint host in the science image) is approximately 20.7 mag \citep{Bellm_2019}. The \textit{r}-band magnitude distribution of the hosts are presented in Fig.~\ref{fig:host_magnitude_distribution}.

We find that 66.31\% of the hosts retrieved by catalogue cross-matching are fainter than the ZTF 20.7 mag limit. This confirms that the majority of the hosts were too faint to be detected in the shallower ZTF data, resulting in them satisfying the conditions to be flagged by ELEPHANT. Some of the remaining cases arise from the intrinsic limitations of the input data rather than failures of the pipeline itself. In most cases, hosts lie outside the 15×15 pixel image cutouts used by the pipeline, in which case a hostless classification is expected, since no off-center information is used. Other cases are affected by poor-quality image stamps, including artefacts, low signal-to-noise regions, or corrupted cutouts 
\citep[see Fig. 3 of][for example]{Pessi_2024}. Finally, a few sources are embedded in hosts that fully occupy the stamp, leaving no clear background region, spatial structure or variability in the shuffled representations to be detected; these are therefore also flagged as hostless.
Note that such candidates are not considered  contaminants in this work. Their hostless identification is consistent with the assumptions of our pipeline and reflects the finite stamp size and data quality constraints, rather than algorithmic shortcomings or transient miss-identification (Section~\ref{sec:contaminants}). We note that cases where the host lies outside the stamp center but contributes some flux within it could be further filtered by adjusting the KS-test thresholds described in Section~\ref{sec:pipeline_recap}.

\subsection{Visual inspection}

\begin{figure}
    \centering
    \includegraphics[width=0.5\textwidth]{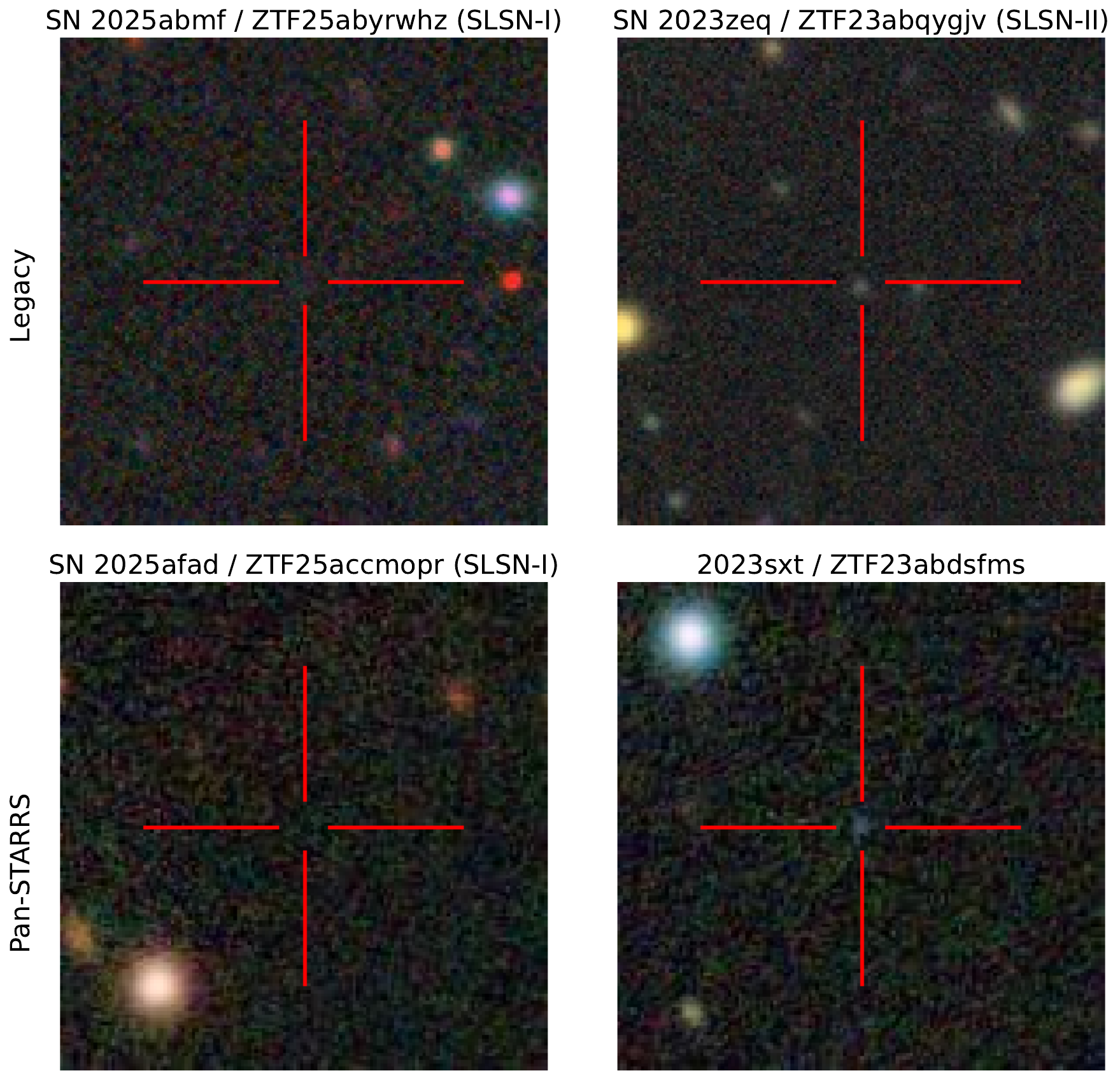}
\caption{Examples of candidates without host association after cross-matching using Sherlock and Pröst. The first and second rows display cutouts from Legacy DR10 and Pan-STARRS DR1 images, respectively. In our visual inspection, we flag cutouts in the left column as hostless and those in the right column as having visually identified host candidates.}
    \label{fig:cutouts_visualization}
\end{figure}

Since some galaxies remain uncatalogued, we performed a visual inspection of the surroundings of the 116 hostless candidates lacking an identified associated host (Fig. \ref{fig:cutouts_visualization}). To do this, we used the LS-DR10 layer and DR9 photoz option in the Legacy Survey Sky Viewer\footnote{\url{https://www.legacysurvey.org/viewer}} to investigate the location of the transient source, searching for faint or previously uncatalogued host galaxies. For Pan-STARRS, we retrieved image cutouts\footnote{\url{https://ps1images.stsci.edu/cgi-bin/ps1cutouts}} at the source's position to perform a similar search. 

We identified plausible host counterparts for 51 candidates. Although we do not perform further analysis of these galaxies, we consider these events to have been correctly flagged as hostless candidates, as the hosts are not visible in the ZTF stamps. The detection of these faint systems supports our catalog-independent approach, which goes beyond cross-matching with existing galaxy catalogs. We expect that such faint galaxies will be visible in LSST stamps, even if they are absent from current catalogs. Moreover, we anticipate that ELEPHANT will be able to reject them without additional information.

\subsection{Truly hostless-candidates}

After removing contaminant sources and excluding all events with a catalogued host galaxy, 65 transients remain as ``truly hostless-candidates'' (Table \ref{tab:true_hostless_candidates}), within the constraints described in this work. Among those, 16 have spectroscopic classification available on TNS.

\begin{table*}
\centering
\renewcommand{\arraystretch}{1.2}
\setlength{\tabcolsep}{3pt}
\begin{tabular}{llllcccc}
\hline\hline
& ZTF Object ID & TNS Name & TNS Classification & Redshift & RA  & DEC & \texttt{Host\_Mag\_r\_upper\_lim} \\
\hline
1 & ZTF25abyrwhz & 2025abmf & SLSN-I & 0.124 & 9:16:12.09 & +49:03:53.46 & -14.03 \\
2 & ZTF23aboebgh & 2023wml & SLSN-I & 0.150 & 11:39:08.69 & -11:14:57.93 & -14.48 \\
3 & ZTF25accmopr & 2025afad & SLSN-I & 0.307 & 8:57:10.72 & -15:33:35.06 & -16.24 \\
4 & ZTF25aaofttb & 2025kkb & SLSN-I & 0.580 & 13:11:37.50 & +10:19:59.22 & -17.88 \\
5 & ZTF25aafwxov & 2025byq & SN Ia & 0.023 & 7:12:27.10 & +50:17:35.73 & -10.21 \\
6 & ZTF23abndxzj & 2023vxt & SN Ia & 0.042 & 23:31:54.34 & +72:50:47.45 & -11.55 \\
7 & ZTF25acdceqb & 2025acsj & SN Ia & 0.050 & 11:33:55.97 & +49:23:02.28 & -11.94 \\
8 & ZTF24aaofopg & 2024jdf & SN Ia & 0.068 & 17:01:22.40 & +78:51:09.51 & -12.64 \\
9 & ZTF23abaslfm & 2023rbt & SN Ia & 0.069 & 1:46:31.43 & +11:51:55.32 & -12.67 \\
10 & ZTF25acaypgx & 2025abzy & SN Ia & 0.070 & 23:12:34.39 & +36:24:47.97 & -12.71 \\
11 & ZTF24abifmph & 2024vwh & SN Ia & 0.110 & 7:00:09.79 & +33:09:58.87 & -13.75 \\
12 & ZTF24aaimwud & 2024lni & SN Ia & 0.115 & 12:41:48.63 & +63:01:06.20 & -13.85 \\
13 & ZTF23abvbwys & 2023aajn & SN Ia-91T-like & 0.049 & 3:41:33.85 & -2:46:50.06 & -11.90 \\
14 & ZTF25abqkwce & 2025xju & SN Ia-SC & 0.040 & 23:55:14.68 & +39:45:49.38 & -11.44 \\
15 & ZTF24aagnlaw & 2024dvi & SN Ia-SC & 0.130 & 10:40:02.35 & +61:50:37.88 & -14.14 \\
16 & ZTF25acgdfwz & 2025agik & SN Ib/c & 0.018 & 1:34:42.98 & +36:57:51.17 & -9.67 \\
\hline
\end{tabular}
\caption{List of truly hostless-candidates with spectroscopic classification. The first column shows the ZTF object name, and the next columns show the TNS name, spectroscopic classification, and corresponding redshift, respectively. \texttt{Host\_Mag\_r\_upper\_lim} is the estimated host galaxy absolute magnitude upper limits, assuming a fixed Legacy Survey \textit{r}-band magnitude of 24.7 mag. The full list of all candidates (including those without spectroscopic classification) is available on GitHub\footref{fn:supplementary_data_ref}.  }
\label{tab:true_hostless_candidates}
\end{table*}

\begin{figure*}
    \centering
    \includegraphics[width=1.\textwidth]{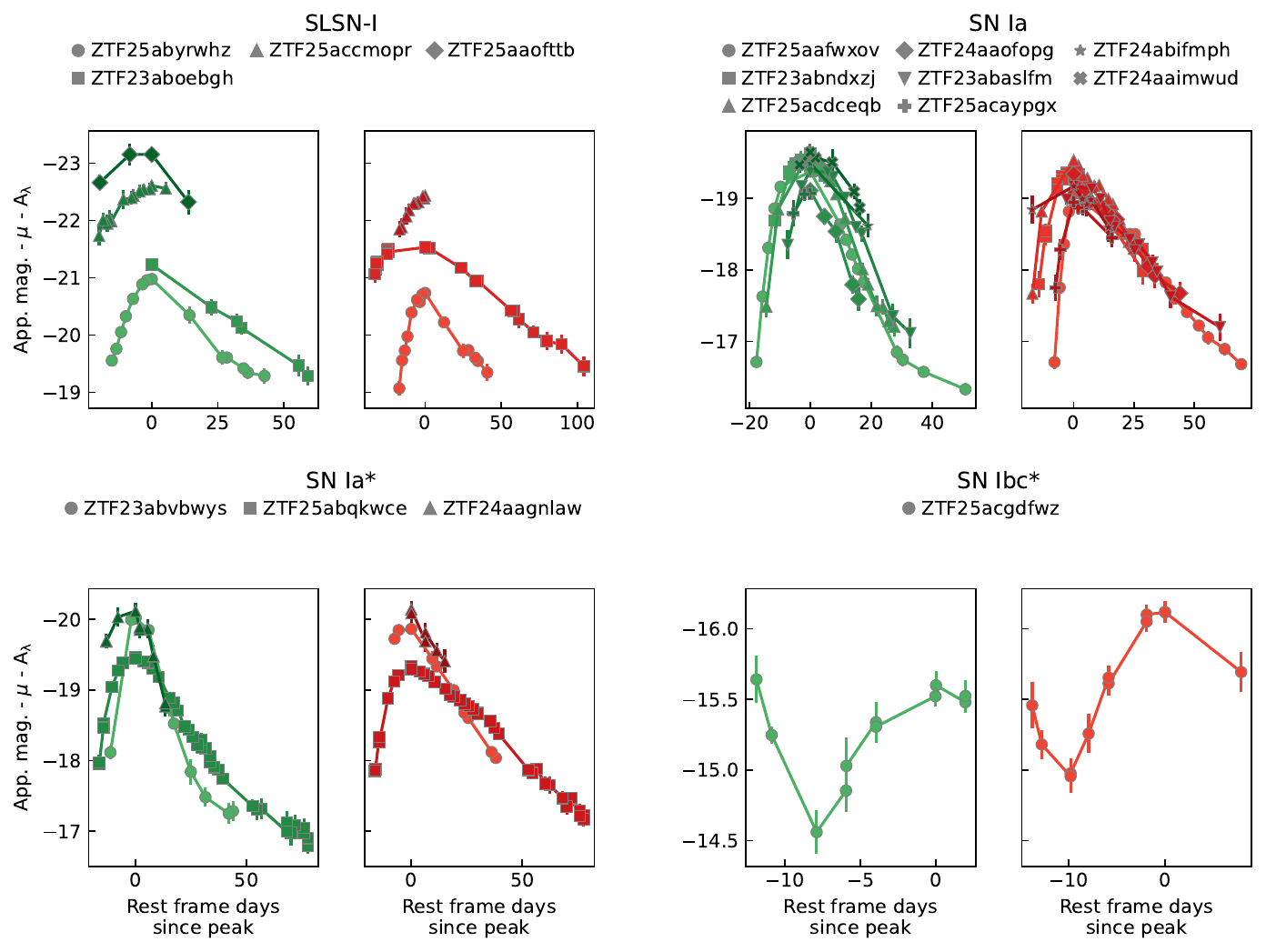}
    \caption{Light curves of truly hostless-candidates with an spectroscopic classification available on TNS. Each subplot shows the $g$ (left, green) and $r$ (right, red) band light curves with respect to the estimated peak. Different events are shown in different symbols and different transparencies. 
    }
    \label{fig:LCs_classified_truly_hostless}
\end{figure*}

Fig.~\ref{fig:LCs_classified_truly_hostless} shows the light curves of the truly hostless-candidates with available spectroscopic classification on TNS, these include 4 SLSN-I (25\%), 8 SN~Ia (50\%), 3 SN~Ia* (18.75\%, encompassing the SN~Ia-91T-like and SN Ia-SC subclasses), and one SN Ib/c. The plot shows phases with respect to the estimated peak, which is considered to be the brighter photometric point in the observed light curve, 
versus $\mathrm{m}_{\lambda} - \mu - \mathrm{A}_{\lambda}$, where $\mathrm{m}_{\lambda}$ is the apparent magnitude retrieved from Fink, $\mu$ is the distance modulus, calculated from the TNS reported redshift using \texttt{astropy.cosmology} software and adopting H$_{0} = 73$ km s$^{-1}$ Mpc$^{-1}$, $\Omega _{\mathrm{Matter}} = 0.27$, $\Omega _{\mathrm{Lambda}} = 0.73$ as cosmological parameters, and $\mathrm{A}_{\lambda}$ is the Milky Way extinction obtained from the NASA/IPAC Extragalactic Database's (NED\footnote{The NASA/IPAC Extragalactic Database (NED) is funded by the National Aeronautics and Space Administration and operated by the California Institute of Technology.}) using NED's Galactic Extinction Calculator\footnote{NED's Extinction Calculator considers the recalibration presented by \citet{2011ApJ...737..103S} to the extinction map presented by \citet{1998ApJ...500..525S}, assuming a \citet{1999PASP..111...63F} reddening law with R$_{\mathrm{v}} = 3.1$.}, accessed through the \texttt{ned\_extinction\_calc} script\footnote{\url{https://github.com/mmechtley/ned_extinction_calc}}. We do not call these absolute magnitudes because we do not consider host extinction or k-corrections, even for the events at higher redshift (see Table \ref{tab:true_hostless_candidates}). 

Below we briefly describe the characteristics of each of the SN classes:
\begin{itemize}
    \item Among the SLSN-I, we see a diversity of light curve shapes and luminosities. SLSN-I are typically associated to low-luminosity, low-metallicity hosts with comparatively high star formation rates \citep[e.g.][]{2014ApJ...787..138L}. Among the flagged events, the most luminous one is SN~2025kkb/ZTF25aaofttb ($-$23.1~mag at peak in the $g$-band), a SLSN-Ic at z=0.58 discovered by \cite{2025TNSTR1783....1P} on 2025-05-14, reported as a hostless candidate by \cite{2025TNSAN.159....1D} on 2025-05-27 and as a  SLSN candidate by \cite{2025TNSAN.179....1L} on 2025-06-11, and finally classified by \cite{2025TNSAN.196....1H} on 2025-06-28.  This event seems to rise faster than the other flagged SLSN-I and may be among the most luminous SLSN-I to date, with other events in the literature peaking slightly below $-$23~mag \citep[e.g.][]{2018ApJ...854...37S,2023ApJ...943...41C,2024MNRAS.535..471G}. 
    \item The SN~Ia light curves show typical behavior and luminosities. Around 37\% of these were classified by the Early SN~Ia module within Fink \citep{2022A&A...663A..13L}, the remainder were not flagged due to gaps or poor sampling of the rising portion of their light curves. The host morphology of SN~Ia, as well as the distance of the SN to the center of the host, are important since they can impact the Hubble diagram residuals \citep{2020MNRAS.499.5121P,2020ApJ...901..143U,2021ARep...65.1015P}, in the case of hostless SN~Ia such parameters cannot be measured, thus it is unclear what is the role of hostless SN~Ia in cosmological studies.
    \item The sample also includes three ``non-normal'' SN~Ia, SN 2025xju/ZTF25abqkwce classified as a SN Ia-SC \citep{2025TNSCR3816....1W}, SN~2024dvi/ZTF24aagnlaw classified as a SN Ia-SC \citep{2024TNSCR.732....1J}, and  SN~2023aajn/ZTF23abvbwys classified as a SN~Ia-91T-like \citep{2023TNSAN.352....1S}. SN~Ia-91T-like are usually found in hosts with ongoing star formation, with many associated to spiral galaxies \citep{2022ApJ...938...47P,2024ApJ...969...80C} while SN~Ia-SC are predominantly found in low-metallicity environments, in the outskirts of their hosts \citep{2011ApJ...737L..24K}. \cite{2011ApJ...733....3C} suggests that low luminosity galaxies host many unusual SN~Ia.
    \item Finally, there is SN~2025agik/ZTF25acgdfwz, classified as a SN~Ib/c by \cite{2026TNSCR.115....1S}. This event shows a pre-peak decline and peaks at $\sim -$16~mag in $r$-band, and at $\sim -$15.5~mag in $g$-band, lying at the fainter end of the peak magnitude distribution of these type of events \citep{2014AJ....147..118R,2018A&A...609A.136T}.
\end{itemize}
 
The upper limits on the host-galaxy absolute magnitudes for truly hostless transients, assuming a fixed \textit{r}-band limiting magnitude of 24.7 mag from the Legacy Surveys, span 
from $M_{r\_upper\_lim} \approx -9.6$ to $-18$, as shown in Table \ref{tab:true_hostless_candidates}. These limits indicate that potential hosts are fainter than normal dwarf galaxies.  Previous studies on hostless transients deem most of the  non-detected host galaxies to be a star-forming dwarf in the field \citep{qin2024statisticsenvironmentshostlesssupernovae}, 
this assumption is based on the low probability of hostless events resulting from hypervelocity stars or arising from intracluster starlight \citep[see also][]{Strolger_2025}. Robustly constraining the physical nature of the host environments for the sample of truly hostless-candidates will require dedicated, deeper follow-up observations. A detailed characterization of such hosts is beyond the scope of this work but represents a crucial step for future studies aiming to fully understand the origin of these transients. 

\begin{figure}
    \centering
    \includegraphics[width=0.5\textwidth]{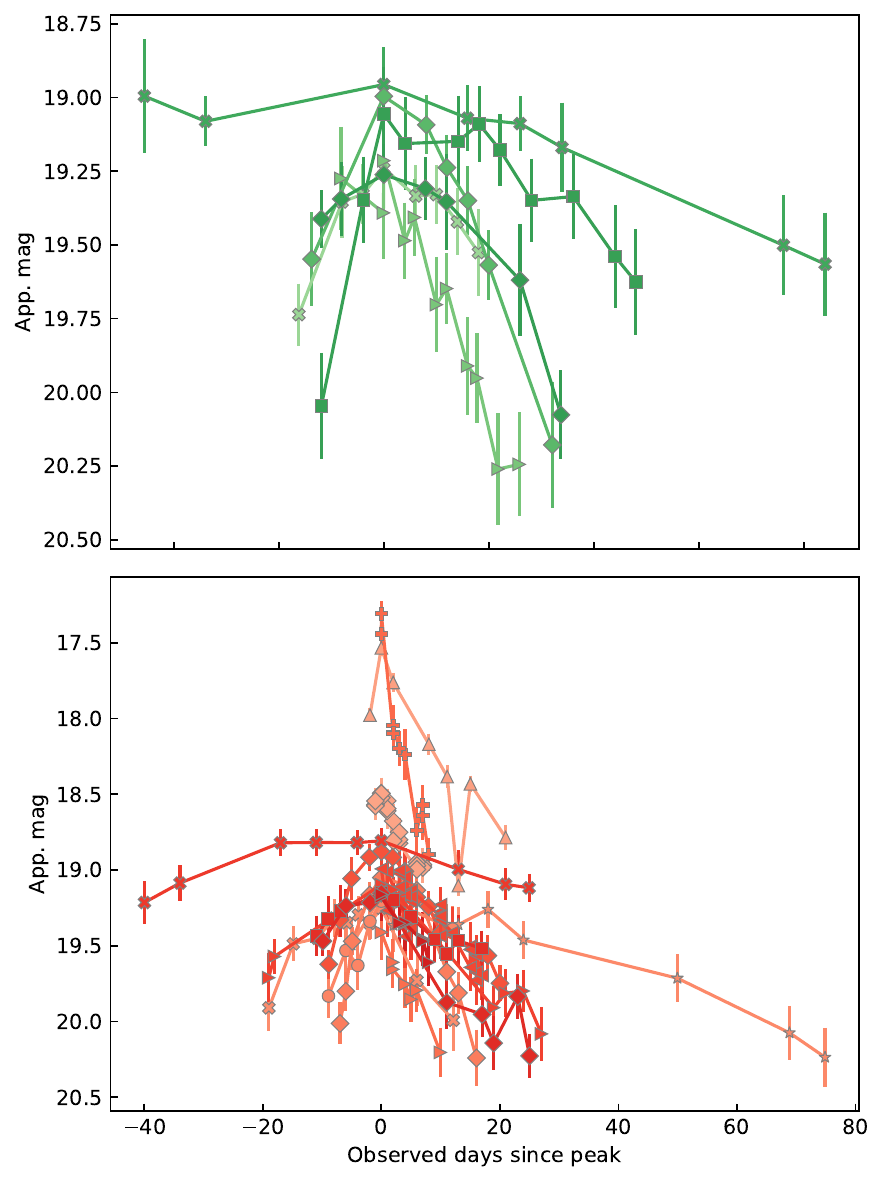}
    \caption{Light curves of truly hostless-candidates without an spectroscopic classification on TNS and with more than six observed photometric points. Each subplot shows the $g$ (top, green) and $r$ (bottom, red) band light curves with respect to the estimated peak. Since there are no available redshifts for these events, we cannot estimate rest frame phases so we report observed ones. Different events are shown in different symbols transparencies. 
    }
    \label{fig:LCs_NOTclassified_truly_hostless}
\end{figure}

Fig.~\ref{fig:LCs_NOTclassified_truly_hostless} shows the light curves of the truly hostless-candidates with no available spectroscopic classification and more than six observed photometric points (the full list of events is presented\footref{fn:supplementary_data_ref}). Most events rise and decline in less than 40 days, with only a few lasting longer. These events cover a wide range of apparent magnitudes, with the brightest one being AT~2024lm/ZTF24aaaejlv, whose  brightest photometric point is at $\sim$ 17 mag, in $g$-band. It was originally reported as a CV candidate by \cite{2024TNSTR..56....1Z}, and identified as a SN candidate by the Fink broker and classified as a SN~Ia or a SN~Ibc by the Alerce broker's\footnote{\url{https://alerce.online/}} \citep{2021AJ....161..242F} light curve classifier \citep{2021AJ....161..141S}. The faintest event is AT~2024sba/ZTF24abalacb, whose brightest photometric point is at $\sim$ 19.9 mag in $g$-band. It was originally reported as a CV candidate by \cite{2024TNSTR2899....1X} and subsequently  identified as a SN candidate by the Fink broker and classified as a SN~Ibc by the Alerce broker's Astronomical Transformer for time series and Tabular data (ATAT) light curve classifier \citep{2024A&A...689A.289C}.

\section{Hostless pipeline for LSST}
\label{sec:elephant_for_lsst}

A significant portion of the difficulties described above, which resulted in the presence of events flagged as hostless candidates even when they had a host association, came from ZTF intrinsic magnitude limitations.  We expect that such effects will be significantly suppressed once the Vera C. Rubin Observatory survey officially starts. Its observations will achieve depths comparable to or exceeding those of existing deep sky surveys with a limiting magnitude for 30-second single \textit{r}-band exposure of $\sim$24.7 mag \citep{bianco2022optimizationcadence}. Even considering that   the limitations associated to the size of the stamps will remain, the detection of a hostless transient in an LSST alert stream will more likely correspond to either the presence of an exceptionally faint host galaxy or a genuinely hostless event. 

In order to ensure that such candidates are identified from the very first alerts produced by Rubin, we have integrated a modified version of the pipeline to the Fink broker. The methodology remains virtually unchanged and leverages additional information available within the LSST alert package and Fink added values for pre‑filtering of alerts.

As new feature, the LSST pipeline incorporates cross-match results from the Gaia DR3 \citep{refId0}, Legacy DR8 \citep{10.1093/mnras/stac608}, SIMBAD \citep{2000A&AS..143....9W}, SPICY \citep{2021AAS...23732901K}, 2MASS \citep{2006AJ....131.1163S}, and HyperLEDA \citep{White:2011qf} hosted by Mangrove. If an alert is associated with any of these catalogs, it is immediately discarded. Alerts with negative \texttt{psfFlux} or a signal-to‐noise ratio for \texttt{templatFlux} $\leq 5$ are filtered out, as well as those located in the galactic plane ($|b| > 20$) and apparent magnitude $\geq 21.5$.

In order to mitigate contamination from SSOs, we initially relied on the \textit{ssObjectId} field provided in the LSST alert packet. However, at early stages of alert production, this identifier can be incomplete or inaccurate. To improve reliability, we implement an additional filtering step based on astrometric behavior: a first-order polynomial is fitted to the right ascension and declination evolution over time. Empirically, SSOs exhibit coherent, approximately linear motion across the sky, resulting in a trajectory that is well described by a straight line, in contrast to the localized positional scatter expected for extragalactic transients. This step effectively removes the majority of residual SSO contaminants. The pipeline flags an object as an SSO if the fitted angular speed is > 2 arcsec/hour and the RMS residual of the fit is < 0.5, and the transient must have at least 3 detections for the fitting to be considered. Furthermore, due to the smaller stamp size in  LSST when compared with ZTF, the Rubin pipeline extracts $30 \times 30$ pixel cutouts centered on the position of each alert (corresponding  to $\sim$ 6 x 6 arcsec). This size is chosen to ensure sufficient spatial context for subsequent analyses, including sigma-clipping and power-spectrum characterization, while remaining computationally efficient within the real-time alert processing framework. 

\begin{figure*}
    \centering
    \includegraphics[scale=0.6]{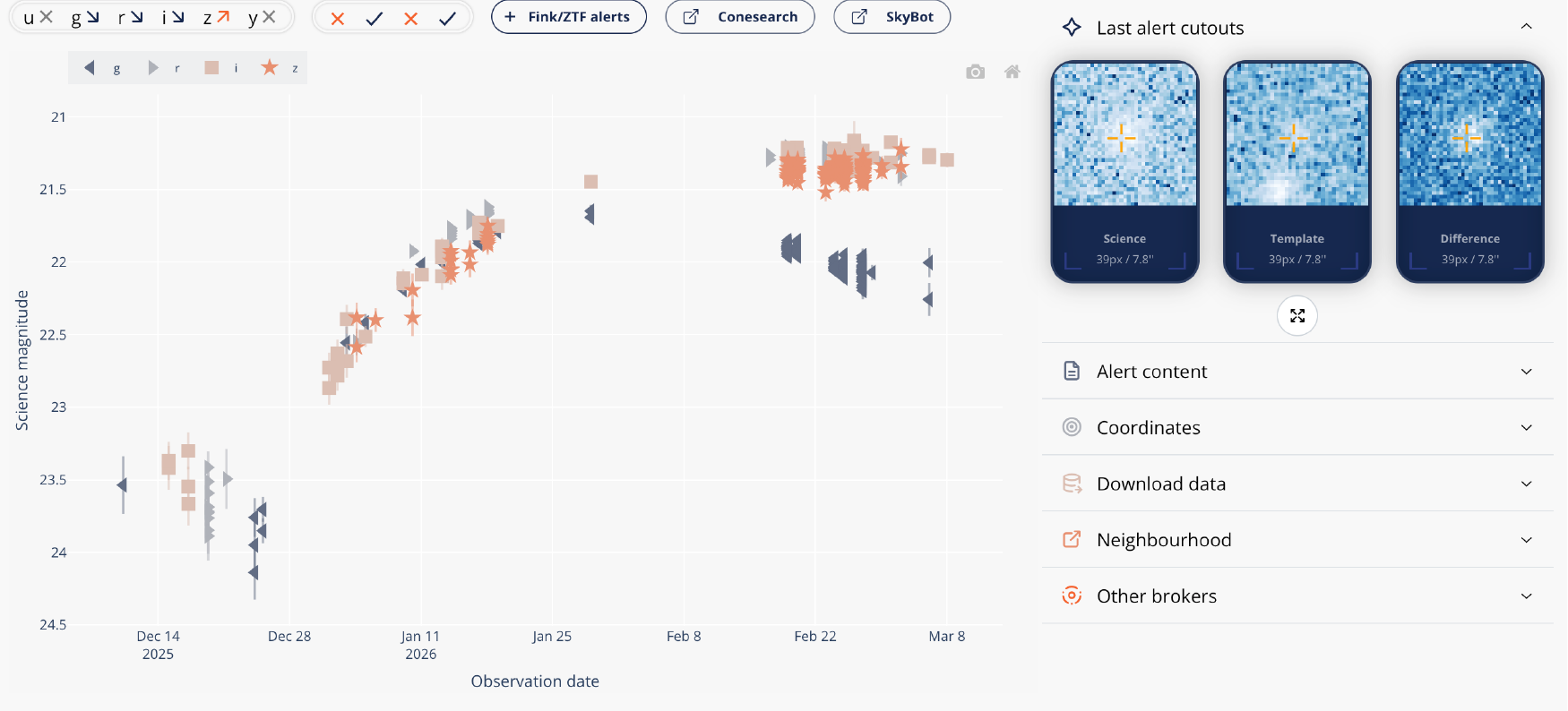}
        \includegraphics[scale=0.6]{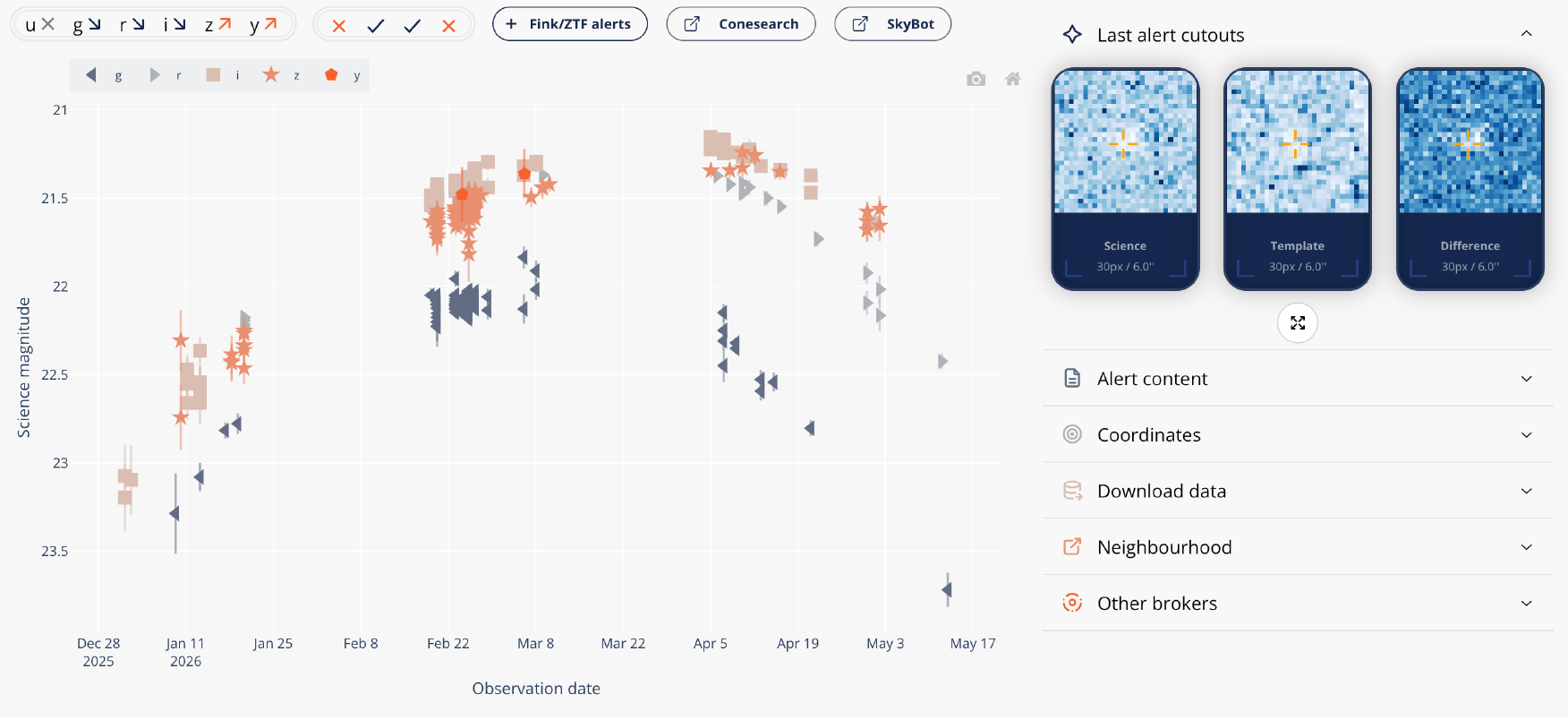}
\caption{Screen shot from the Fink Rubin Portal, showing two of the reported Rubin transients from our pipeline: the top panel shows 313831458352922664\protect\footnotemark and  the bottom one,  313928194567700757 \protect\footnotemark, both of which have been spectroscopically classified as SLSN-I.}
 \label{fig:rubin_reported_candidates}
\end{figure*}

We have tested this adaptation on the initial LSST alert stream. Three candidates were found and reported, being among the first set of LSST alerts on TNS \citep{2026TNSAN..49....1D}, two of them (Figure \ref{fig:rubin_reported_candidates}) have been later classified spectroscopically:  SN 2026ejf \citep[SLSN-I,][]{2026TNSAN..72....1V} and  SN 2025altn \citep[SLSN-I,][]{2026TNSCR1048....1J}. The third one, AT~2025altt has been associated to a SN~IIn by the Superphot+ light curve classifier \citep{2024ApJ...974..169D,2026TNSAN..50....1S}. Future developments will certainly include cross-matching with Rubin specific source catalogs and host association techniques \citep[e.g.][]{weston2026}.

 \addtocounter{footnote}{-1}
\footnotetext{\url{https://lsst.fink-portal.org/313831458352922664}}
\stepcounter{footnote}
\footnotetext{\url{https://lsst.fink-portal.org/313928194567700757}}

\section{Conclusion}
\label{sec:conclusions}

We analyzed the performance of the hostless pipeline implemented in the Fink alert broker using over two years of ZTF alerts. The pipeline identified 877 hostless candidates within a total number of 156 110 unique events observed in the same period, of which 276 have spectroscopic classifications in TNS. Most of the flagged transients have been classified as Type Ia supernovae (162); many others (30) have been identified as Type I superluminous supernovae, while another set (48) belong to other supernova sub-classes. Thus, our major source of contamination are CVs, along with 3 moving stars. Under the assumption that only non-extragalactic and variable stellar sources constitute contaminants, the pipeline achieves an overall accuracy of $\sim$84\%.

After cross-matching with deeper sky survey catalogues, using  Sherlock and Pröst, we find that 116 of the flagged hostless transients lack catalogued host associations. We note that 25\% of the identified hosts show a magnitude discrepancies higher than 0.5~mag in the $r$-band when the results of both packages are compared, suggesting potential inconsistencies in host identification.

We find that 33.69\% of events are associated to hosts brighter than the ZTF detection limit; in most cases, these hosts fall outside the ZTF stamp boundaries. These are not considered contaminants, since the portion of the stamp analyzed by the pipeline fulfills our criteria for hostless characterization. We note that adjusting the KS-test thresholds could modify this fraction.

Upon visual inspection of the Legacy Survey and Pan-STAARS images, we identified 65 transients as truly hostless-candidates, while the remainder 51 have a source that is visually detected on these catalogues; among them, 16 have spectroscopic classifications in TNS. The inferred upper limits of  host-galaxy absolute magnitude for these range from $M_{r,\mathrm{lim}} \approx -9.6$ to $-18$, consistent with extremely faint or low-mass host systems.

We have updated the pipeline to mitigate contamination from moving stars, uncatalogued asteroids and to enable real time processing of the LSST alert stream within the Fink broker. While the pipeline is effective at selecting hostless transient candidates (only $\sim 0.6 \%$ of the events in the analyzed period were flagged as hostless candidates), manual visual inspection is still necessary to confirm the most interesting cases and avoid misclassifications. Following the its natural development, this will certainly lead to improvements culminating in a reliable sample of LSST hostless extragalactic transients to the community. At that point, a change of paradigm will be unavoidable. 

If until now we have been able to assign the hostless aspect of many extragalactic transients to the limiting magnitude of large scale sky surveys, the situation will drastically change with the construction of a coherent sample of Rubin candidates. Considering that arguments for the low incidence of supernovae from runaway stars will continue to hold, and that the larger depth of LSST alert stamps will remove most of the current limitations in identifying low surface brightness galaxies, new theoretical frameworks will be required to explain the number and diversity of the surviving hostless candidates. Thus fomenting theoretical developments in stellar and galaxy formation and dynamics. The work presented here was designed to sparkle that revolution.

\begin{acknowledgements}
P.J.P acknowledges funding from the European Union’s Horizon Europe 914
Research and Innovation Programme under Grant Agreement No. 101131928 915
(ACME). 
This work was developed within the Fink community and made use of the Fink resources. Fink is supported by LSST-France and CNRS/IN2P3. This work is based on observations obtained with the Samuel Oschin Telescope 48-inch and the 60-inch Telescope at the Palomar Observatory as part of the Zwicky Transient Facility project. ZTF is supported by the National Science Foundation under Grants No. AST-1440341 and AST-2034437 and a collaboration including current partners Caltech, IPAC, the Weizmann Institute of Science, the Oskar Klein Center at Stockholm University, the University of Maryland, Deutsches Elektronen-Synchrotron and Humboldt University, the TANGO Consortium of Taiwan, the University of Wisconsin at Milwaukee, Trinity College Dublin, Lawrence Livermore National Laboratories, IN2P3, University of Warwick, Ruhr University Bochum, Northwestern University and former partners the University of Washington, Los Alamos National Laboratories, and Lawrence Berkeley National Laboratories. Operations are conducted by COO, IPAC, and UW.
This research has made use of the NASA/IPAC Extragalactic Database, which is funded by the National Aeronautics and Space Administration and operated by the California Institute of Technology.
This research has made use of the SVO Filter Profile Service "Carlos Rodrigo", funded by MCIN/AEI/10.13039/501100011033/ through grant PID2020-112949GB-I00
The Legacy Surveys consist of three individual and complementary projects: the Dark Energy Camera Legacy Survey (DECaLS; Proposal ID 2014B-0404; PIs: David Schlegel and Arjun Dey), the Beijing-Arizona Sky Survey (BASS; NOAO Prop. ID 2015A-0801; PIs: Zhou Xu and Xiaohui Fan), and the Mayall z-band Legacy Survey (MzLS; Prop. ID 2016A-0453; PI: Arjun Dey). DECaLS, BASS and MzLS together include data obtained, respectively, at the Blanco telescope, Cerro Tololo Inter-American Observatory, NSF’s NOIRLab; the Bok telescope, Steward Observatory, University of Arizona; and the Mayall telescope, Kitt Peak National Observatory, NOIRLab. Pipeline processing and analyses of the data were supported by NOIRLab and the Lawrence Berkeley National Laboratory (LBNL). The Legacy Surveys project is honored to be permitted to conduct astronomical research on Iolkam Du’ag (Kitt Peak), a mountain with particular significance to the Tohono O’odham Nation.

NOIRLab is operated by the Association of Universities for Research in Astronomy (AURA) under a cooperative agreement with the National Science Foundation. LBNL is managed by the Regents of the University of California under contract to the U.S. Department of Energy.

This project used data obtained with the Dark Energy Camera (DECam), which was constructed by the Dark Energy Survey (DES) collaboration. Funding for the DES Projects has been provided by the U.S. Department of Energy, the U.S. National Science Foundation, the Ministry of Science and Education of Spain, the Science and Technology Facilities Council of the United Kingdom, the Higher Education Funding Council for England, the National Center for Supercomputing Applications at the University of Illinois at Urbana-Champaign, the Kavli Institute of Cosmological Physics at the University of Chicago, Center for Cosmology and Astro-Particle Physics at the Ohio State University, the Mitchell Institute for Fundamental Physics and Astronomy at Texas A\&M University, Financiadora de Estudos e Projetos, Fundacao Carlos Chagas Filho de Amparo, Financiadora de Estudos e Projetos, Fundacao Carlos Chagas Filho de Amparo a Pesquisa do Estado do Rio de Janeiro, Conselho Nacional de Desenvolvimento Cientifico e Tecnologico and the Ministerio da Ciencia, Tecnologia e Inovacao, the Deutsche Forschungsgemeinschaft and the Collaborating Institutions in the Dark Energy Survey. The Collaborating Institutions are Argonne National Laboratory, the University of California at Santa Cruz, the University of Cambridge, Centro de Investigaciones Energeticas, Medioambientales y Tecnologicas-Madrid, the University of Chicago, University College London, the DES-Brazil Consortium, the University of Edinburgh, the Eidgenossische Technische Hochschule (ETH) Zurich, Fermi National Accelerator Laboratory, the University of Illinois at Urbana-Champaign, the Institut de Ciencies de l’Espai (IEEC/CSIC), the Institut de Fisica d’Altes Energies, Lawrence Berkeley National Laboratory, the Ludwig Maximilians Universitat Munchen and the associated Excellence Cluster Universe, the University of Michigan, NSF’s NOIRLab, the University of Nottingham, the Ohio State University, the University of Pennsylvania, the University of Portsmouth, SLAC National Accelerator Laboratory, Stanford University, the University of Sussex, and Texas A\&M University.

BASS is a key project of the Telescope Access Program (TAP), which has been funded by the National Astronomical Observatories of China, the Chinese Academy of Sciences (the Strategic Priority Research Program “The Emergence of Cosmological Structures” Grant XDB09000000), and the Special Fund for Astronomy from the Ministry of Finance. The BASS is also supported by the External Cooperation Program of Chinese Academy of Sciences (Grant 114A11KYSB20160057), and Chinese National Natural Science Foundation (Grant 12120101003, 11433005).

\end{acknowledgements}

\bibliography{analysis_paper/main.bib}

\appendix
\section{Fink alert filters}\label{app:fink_filters}

The ELEPHANT pipeline processes alerts that have any of the following classifications, which are added within the Fink broker. Fink classifiers are added using machine learning or statistical algorithms. TNS and SIMBAD classifications are provided by the broker through cross‑matching with those catalogues.
\begin{itemize}
    \item Fink classifiers:
    Early SN Ia candidate, SN candidate, Kilonova candidate.

    \item TNS classifiers:
    Afterglow, AGN, Galaxy, FRB, Impostor-SN, ILRT,
    Kilonova, LBV, Light-Echo, LRN, Nova, Other, QSO, SLSN-II, SN, SN I, SN Ia,
    SN Ia-91bg-like, SLSN-I, SN Ia-91T-like, SN Ia-CSM, SN Ia-pec,
    SN Iax[02cx-like], SN Ib, SN Ib-Ca-rich, SN Ib-pec, SN Ib/c, SN Ibn,
    SN Ic, SN Ic-BL, SN Ic-pec, SN Icn, SN II, SN II-pec, SN IIb,
    SN IIL, SN IIn, SN IIn-pec, SN IIP, TDE.

    \item SIMBAD classifiers:
    AGN, \texttt{AGN\_Candidate}, \texttt{BH\_Candidate}, Blazar,
    \texttt{Blazar\_Candidate}, BLLac, \texttt{BLLac\_Candidate},
    \texttt{Candidate\_Nova}, \texttt{Candidate\_NS}, \texttt{Candidate\_SN*},
    LINER, QSO, \texttt{QSO\_candidate}, Seyfert, Seyfert1, Seyfert2,
    \texttt{Seyfert\_1}, \texttt{Seyfert\_2}, SN,
    \texttt{SN*\_Candidate}, Supernova, ULX, \texttt{ULX?},
    \texttt{ULX\_candidate}.
\end{itemize}

\FloatBarrier 
\clearpage

\end{document}